\documentclass[5p,times]{elsarticle}
\usepackage{amsmath,amssymb}
\usepackage{graphicx}
\graphicspath{{figures/}{./}}
\usepackage{booktabs}
\usepackage{array}
\usepackage{algorithm}
\usepackage{algpseudocode}
\usepackage{url}
\usepackage[hidelinks]{hyperref}
\hypersetup{pdfauthor={Prashant Kumar Pathak, Tarun Kumar Sharma},
            pdftitle={When Global Gating Is Enough: Admission-Time Hubness Control in Anisotropic Vector Retrieval Systems}}
\usepackage{microtype}

\journal{Computers \& Security}

\begin{document}

\begin{frontmatter}

\title{When Global Gating Is Enough: Admission-Time Hubness Control in Anisotropic Vector Retrieval Systems}

\author{%
\begin{tabular}[t]{@{}c@{\hspace{3em}}c@{}}
\textbf{Prashant Kumar Pathak} & \textbf{Tarun Kumar Sharma} \\[3pt]
\textit{Santa Clara, CA, USA} & \textit{Colonia, NJ, USA} \\[2pt]
\texttt{prashant.pathak@ieee.org} & \texttt{tarun.sharma@ieee.org}
\end{tabular}}

\begin{abstract}
Vector \emph{hubness}---the tendency of a few points to be nearest neighbours of disproportionately many
queries in high-dimensional space---is an attack surface for retrieval-augmented generation (RAG): an
adversary injects one crafted document that is retrieved for many unrelated queries, turning a single
record into a broad poisoning vector. The prevailing defence is \emph{detection}: periodically scan the
index, estimate reverse-$k$NN influence, and remove outliers, which leaves an exposure window and
requires repeated corpus-wide rescans. We study \emph{admission-time} control---scoring each document's
hub behaviour against sentinel queries and refusing it before it becomes retrievable---and the systems
question of whether the control can be maintained \emph{incrementally} rather than by rescans. On a
$100{,}000$-document corpus assembled from four BEIR collections ($500$ topics, $5{,}571$ sentinels,
\texttt{bge-large-en-v1.5}, exact $k$NN) and in an evasion setting where attacker anchors are
\emph{disjoint} from the defender's sentinels, a single \emph{global} admission gate recovers
effective adversarial hubs on embedding-space attacks at recall $1.0$ at the decisive operating point ($\ge 0.92$ across the effective range) and AUROC $\approx 1.0$ at an $\approx1\%$ false-positive rate on in-domain general documents (organic natural hubs remain a residual requiring provenance-based adjudication); on harder gradient-realised
(HotFlip) attacks the global gate averages $0.91\pm0.07$ recall across five seeds, at least as good as a
per-topic gate ($0.85\pm0.05$). The gate matches a recent reverse-$k$NN detector and dominates a
provenance-only baseline that would require trusting $99\%$ of sources. A more elaborate
\emph{domain-aware} (per-topic) gate provides no statistically reliable benefit: its marginal over the global gate is $+0.000$ in the cross-encoder (five encoders, $384$--$1024$ dimensions: MiniLM, BGE, GTE, E5) and cross-corpus (two compositionally distinct $100$k-document corpora) evaluations, while the five-seed suppression sweep yields $+0.018\pm0.025$ (a confidence interval including zero). We give a geometric account: embedding spaces are anisotropic, so per-topic
query centroids are positively aligned with the global centroid (strongly in four of five encoders), so a hub cannot easily be simultaneously topic-loud and globally-quiet; a direct vector-space optimisation over the unit sphere did not find such a vector in the evaluated searches. On the systems
side we maintain the admission thresholds incrementally---exact against full recomputation, with
per-write cost independent of corpus size (amortised for deletion)---and evaluate the full pipeline on a real approximate
(HNSW) index: admission adds $\approx3.1\%$ to ingestion latency ($0.6\%$ for candidate scoring), its candidate-scoring cost is flat to $N=10^6$,
it tolerates approximate indexing ($1.2\%$ of admit/quarantine decisions flip under HNSW, none of them attacks), and concurrent
ingestion scales $5.9\times$ across eight threads. The contribution is a preventive,
incrementally-maintainable \emph{global} admission gate, a reproducible negative result on domain-aware
refinement, and a geometric explanation of when global control suffices; we treat provenance as a
complementary control for the residual (weak or tight-domain hubs).
\end{abstract}


\end{frontmatter}

\section{Introduction}\label{sec:intro}
Retrieval-augmented generation (RAG)~\citep{lewis2020rag,guu2020realm} grounds a language model on
documents fetched from a vector store by embedding similarity. The retrieved passages are treated as
trusted context, so the store is a security boundary: a document retrieved for a query can steer the
model's output for that query. \emph{Hubness}---the concentration, in high-dimensional spaces, of
reverse-nearest-neighbour (reverse-$k$NN) counts on a few points~\citep{radovanovic2010hubs}---turns this into an attack
surface. An adversary who crafts a \emph{hub} document can have it retrieved across many unrelated
queries, so a single injected record influences a large fraction of
interactions~\citep{zhang2024advhub,zou2024poisonedrag,zhong2023corpuspoison}.

The prevailing defence is detection: periodically scan the index, estimate reverse-$k$NN influence, and
remove statistical outliers~\citep{cisco2026ahd}. Detection after the fact has two structural costs: an
\emph{exposure window} between a hub's insertion and the next scan, during which it hijacks queries, and
the expense of rescanning a large store. This paper asks whether the same statistic can be applied
\emph{at admission}---evaluated on each document \emph{before} it becomes retrievable, so a hub never
enters the index---and whether that control can be maintained \emph{incrementally} per write rather than
by repeated full rescans. A useful gate must also separate adversarial hubs from \emph{legitimately}
general documents (glossaries, overviews) that are naturally hub-like, or it would break normal ingestion.

A natural refinement is to condition the statistic on topic (``domain-aware'' detection), on the
intuition that a concept-specific hub may be loud within one topic yet quiet globally and thus slip past
a global threshold. Our central empirical finding is that, on dense single-vector retrieval, this
refinement buys nothing: across the evaluated encoders, corpora, attacks, and operating points, domain-aware gating provides no statistically reliable benefit. We explain this geometrically and show that the global gate is
cheap enough to run synchronously on the ingestion path of a real approximate-nearest-neighbour (ANN) index.

\paragraph{Contributions}
\begin{enumerate}\setlength{\itemsep}{2pt}
\item A \emph{preventive} admission-time formulation of hubness control that evaluates documents before
they are retrievable, removing the exposure window left by detective scanning (Sections~\ref{sec:gate},~\ref{sec:sys}).
\item A \emph{scalable incremental-maintenance} scheme for the admission thresholds---exact against full
recomputation, with per-write cost independent of corpus size (amortised for deletion)---and a real-index systems evaluation
(HNSW; end-to-end latency, scalability to $N=10^6$, approximate-index robustness, concurrency)
showing the gate adds negligible relative overhead in the evaluated ingestion pipeline (Sections~\ref{sec:sys},~\ref{sec:systems}).
\item A \emph{reproducible negative result}: a domain-aware per-topic gate provides no measurable benefit
over the global gate across five encoders ($384$--$1024$\,d; MiniLM/BGE/GTE/E5), two compositionally
distinct corpora, two concept definitions, and a five-seed sweep (Sections~\ref{sec:sec},~\ref{sec:repro}).
\item A \emph{geometric explanation}---embedding anisotropy couples local and global visibility---that
provides a geometric account of the observed sufficiency of global gating and identifies a tight-domain exception
(Sections~\ref{sec:geo},~\ref{sec:residual}).
\end{enumerate}

\section{Threat Model and Assumptions}\label{sec:threat}
\textbf{System.} A vector store holds unit-normalised embeddings of single-passage documents under a
fixed encoder; queries retrieve the top-$k$ by cosine similarity (equivalently inner product on
normalised vectors). \textbf{Adversary.} The attacker injects documents through the ingestion path but
cannot alter the encoder, other stored vectors, or the retrieval code. The goal is a hub retrieved for
many \emph{anchor} queries the attacker cares about. We grant the attacker knowledge of the defence and
the ability to optimise hubs by gradient methods, and---as the realistic evasion case---hold the
attacker's anchor queries \emph{disjoint} from the defender's sentinels (measured overlap $0.000$).
\textbf{Out of scope.} Targeted single-query poisoning~\citep{zou2024poisonedrag,bentov2025gaslite},
multi-vector / late-interaction retrieval~\citep{khattab2020colbert}, and model-internal poisoning are
out of scope and revisited as limitations.

\section{Admission-Time Global Gate}\label{sec:gate}
Let the corpus be $C=\{x_1,\dots,x_N\}$, $x_i\in\mathbb{R}^d$, $\lVert x_i\rVert=1$, and let
$Q=\{q_1,\dots,q_M\}$ be sentinel-query embeddings. For each sentinel $q_j$ define a threshold $\tau_j$
as the similarity of its $k$-th nearest neighbour in $C$, so a candidate $d$ would enter $q_j$'s top-$k$
iff $\langle d,q_j\rangle > \tau_j$. The \emph{global hub score} is the fraction of sentinels the
candidate would displace into,
\begin{equation}
h(d) \;=\; \frac{1}{M}\sum_{j=1}^{M}\mathbf{1}\!\left[\langle d,q_j\rangle > \tau_j\right],
\end{equation}
and the gate admits $d$ if $h(d)\le\theta$, quarantining it otherwise, with $\theta$ calibrated to a
target false-positive budget on held-out natural documents. A normal single-topic document beats only a
small fraction of sentinels; being retrieved by many is the adversarial signal.

\section{Domain-Aware Alternative and Geometric Hypothesis}\label{sec:da}
The natural refinement conditions on topic. With per-topic local hub rate $s_t(d)$ over topic $t$'s
sentinels, a worst-topic robust-$z$ score
\[
z(d)=\max_t \frac{s_t(d)-\mathrm{med}_t}{\max\!\big(1.4826\,\mathrm{MAD}_t,\;1/n_t\big)}
\]
(here $\mathrm{MAD}_t$ is the median absolute deviation; the scale floored at the rate resolution $1/n_t$ avoids a degenerate zero-MAD blow-up) is intended to
catch a hub that is quiet globally but loud within one topic. The \emph{geometric hypothesis} we test is
that, under anisotropy, no such hub exists, so $z(d)$ carries no information beyond $h(d)$
(Section~\ref{sec:geo}).

\section{Incremental Systems Architecture}\label{sec:sys}
The gate sits synchronously on the ingestion path: admitted documents are indexed, flagged documents
quarantined for review. An external, slowly-updated baseline supplies $\theta$ and resists a dilution
attack (mass injection that shifts the corpus's own statistics); a velocity monitor watches for many
sub-threshold documents converging on one region; a periodic detective scan remains as a backstop for
queries outside the sentinel set. Each sentinel maintains its current top-$B$ similarities ($B=k+c$) in a
buffer with row ids, so $\tau_j$ is the $k$-th entry and scoring is one pass over the sentinels
(Algorithms~\ref{alg:score}--\ref{alg:upd}).

\begin{algorithm}[t]
\caption{Candidate scoring and admission}\label{alg:score}
\begin{algorithmic}[1]
\Require candidate $d$, sentinels $Q$, thresholds $\{\tau_j\}$, budget $\theta$
\State $h \gets \frac{1}{M}\sum_{j}\mathbf{1}[\langle d,q_j\rangle > \tau_j]$
\If{$h \le \theta$} \State \Return \textsc{admit} \Else{} \State \Return \textsc{quarantine} \EndIf
\end{algorithmic}
\end{algorithm}

\begin{algorithm}[t]
\caption{Incremental threshold maintenance (insert / delete)}\label{alg:upd}
\begin{algorithmic}[1]
\Procedure{Insert}{$d$}
  \For{$j$ with $\langle d,q_j\rangle > \min B_j$} splice $\langle d,q_j\rangle$ into $B_j$, keep top-$B$; $\tau_j\gets B_j[k]$
  \EndFor
\EndProcedure
\Procedure{Delete}{$x$}
  \For{$j$ such that $x \in B_j$ (scan, $O(MB)$ total)} drop $x$ from $B_j$
     \If{$|B_j| < k$} refill $B_j$ by one rescan \Comment{rare} \Else{} $\tau_j\gets B_j[k]$ \EndIf
  \EndFor
\EndProcedure
\end{algorithmic}
\end{algorithm}

\begin{figure}[t]\centering
\includegraphics[width=0.78\linewidth]{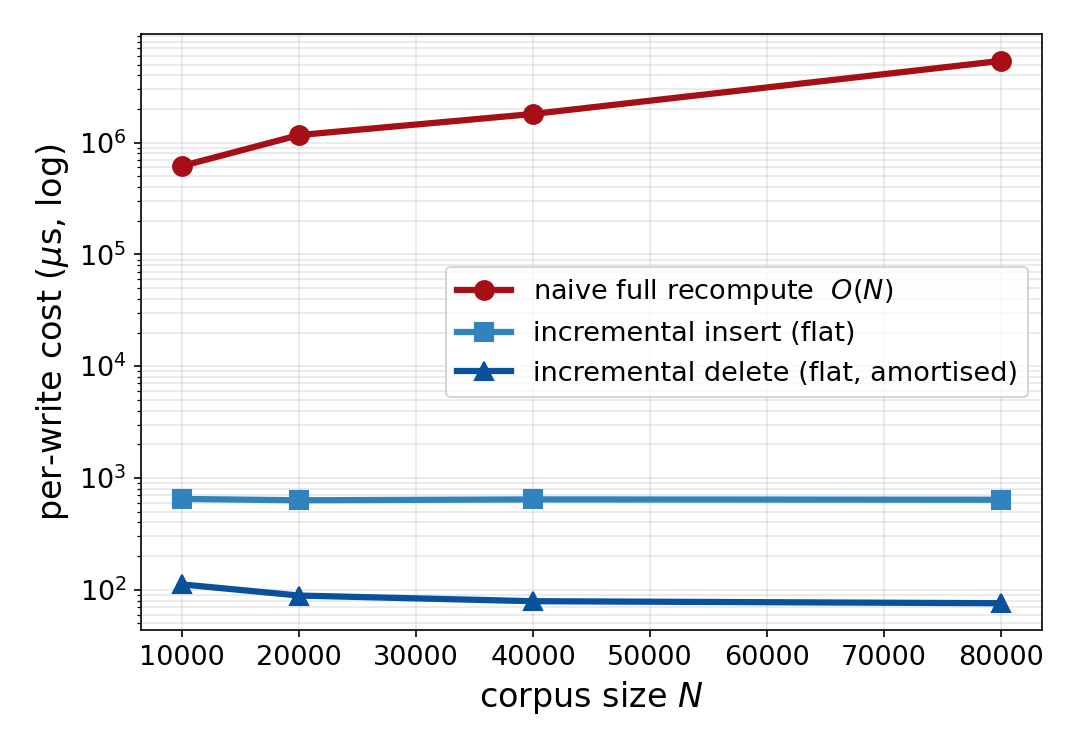}
\caption{Incremental threshold maintenance on the real \texttt{bge-large} embeddings: per-write
insert/delete cost is flat in $N$ and exact against full recomputation, whereas the naive recompute is
$O(N)$.}\label{fig:incr}
\end{figure}

\textbf{Complexity.} Scoring and insertion cost $O(Md)$ and $O(Md + a\log k)$ ($a$ = displaced
sentinels); buffered deletion costs $O(MB)$ for the membership check with an occasional $O(N)$ refill
that amortises to a constant (a sentinel underflows only every $\sim(B-k)$ buffer hits, and the
buffer-hit rate is $\sim 1/N$). Insertions and ordinary deletions are $N$-independent; deletion is amortised $N$-independent under the stated random-delete workload, with an $O(N)$ worst-case refill. Per-write costs grow
with the sentinel count $M$ and embedding dimension $d$ (and, under an approximate index, with the ANN
query used to maintain $\tau$). Space is $O(MB+Md)$. The maintained thresholds are \emph{exact}: on the
real \texttt{bge-large} embeddings at the \emph{deployed} sentinel count ($M{=}5{,}571$, $k{=}10$, $B{=}50$, $200$ randomly generated update operations at each tested corpus size) the per-write update is flat in $N$ ($\approx\!0.72$\,ms insert, $\approx\!0.08$--$0.11$\,ms delete from $N=10^4$ to $8\times10^4$) with maximum error $0$ and zero buffer refills, while a naive full recompute grows linearly ($0.56$\,s$\rightarrow$$4.2$\,s); see Fig.~\ref{fig:incr}. Section~\ref{sec:systems}
evaluates the gate end-to-end on a real approximate index.

\section{Experimental Methodology}\label{sec:method}
\textbf{Corpus.} $N=100{,}000$ documents from four BEIR collections~\citep{thakur2021beir},
gold-prioritised then subsampled: FiQA (finance; the concept target; $57{,}318$), TREC-COVID
($33{,}866$), SciFact ($5{,}183$), NFCorpus ($3{,}633$); passages under $80$ characters dropped;
$10{,}200$ grounded queries; $328/500$ topics contain a query.
\textbf{Encoder.} \texttt{BAAI/\allowbreak bge-\allowbreak large-\allowbreak en-\allowbreak v1.5} (dim $1024$)~\citep{xiao2023bge} via
\texttt{sentence-\allowbreak transformers}~\citep{reimers2019sbert}; \texttt{float32}, \texttt{max\_\allowbreak seq\_\allowbreak length}
$256$, L2-normalised once (cosine $=$ inner product); the query-side instruction is prepended to queries
only. Hardware: Apple-silicon (arm64) with the MPS backend.
\textbf{Topics.} $T=500$ clusters from \texttt{MiniBatchKMeans} on the L2-normalised
embeddings; a query's topic is the majority topic of its gold documents.
\textbf{Search.} Exact brute-force $k$NN ($k=10$) for the separation study; the systems study
(Section~\ref{sec:systems}) uses a real Hierarchical Navigable Small World (HNSW) index~\citep{malkov2018hnsw}. $\tau_j$ is each sentinel's
$k$-th-NN similarity over the clean corpus.
\textbf{Sentinels.} $M=5{,}571$: held-out evaluation queries (half of each topic's queries) plus the
$500$ cluster centroids.
\textbf{Splits.} Within each topic, queries are split evenly into sentinels and attacker anchors;
sentinel--anchor overlap $0.000$.
\textbf{Negatives.} The headline natural-negative class is N-organic $\cup$ N-curated: $1{,}000$
\emph{organic} natural hubs (top-$1\%$ by reverse-$k$-occurrence over the queries) and $1{,}500$
\emph{curated} in-domain definitional/overview passages mined from the corpus (excluding organic). Two
sanity floors: $1{,}500$ cross-domain Simple-English-Wikipedia passages and $2{,}000$ random documents.
A disjoint benign calibration set ($5{,}000$ random documents) supplies the robust-$z$ baseline; attacks
and evaluation negatives are never used for calibration.
\textbf{Calibration.} The operating threshold $\theta$ is selected to give a $1\%$ false-positive rate (FPR) on the \emph{disjoint} $5{,}000$-document benign calibration set (never the evaluation negatives), then frozen; we report attack recall and the per-population false-positive rate at that fixed threshold (Table~\ref{tab:fpr}). For corpus-derived benign candidates (organic, curated, random, calibration), admission scores use leave-one-out thresholds, so each document is evaluated as a pre-insertion candidate; external candidates (injected attacks, cross-domain Wikipedia) use the standard threshold.
\textbf{Attacks.} Universal hubs ($24$, random anchor subsamples); concept hubs by a gradient hub-rate
search on the unit sphere across a suppression sweep $\lambda\in\{0,0.5,1,2,4,8,16\}$ (Table~\ref{tab:lambda} lists a representative subset; $8$ targets, $5$
seeds; an effective concept hub retrieves $\ge 20\%$ of its target topic), with a closed-form mean-cosine
attack as reference; and HotFlip-through-BGE~\citep{ebrahimi2018hotflip} text-constrained hubs ($4$
targets $\times$ $5$ seeds, length $32$, $120$ steps), realising the $\lambda^{*}=2$ direction as valid
but non-fluent token sequences. Cisco's open detector~\citep{cisco2026ahd} (the \texttt{hubscan} software~\citep{cisco2026soft}) is run as a baseline on the same corpus, sentinels, and hubs.
\textbf{Reproducibility.} Five seeds $\{20260612,7,42,123,2024\}$ reuse the cached corpus embedding and re-randomise everything downstream that drives the verdict.
\textbf{Software.} Python 3.11 with \texttt{torch} 2.8, \texttt{sentence-transformers} 5.2, \texttt{scikit-learn} 1.7, \texttt{hnswlib} 0.8, \texttt{numpy}/\texttt{scipy}, on arm64 macOS (MPS); encoder model cards: \texttt{BAAI/bge-large-en-v1.5}, \texttt{BAAI/bge-base-en-v1.5}, \texttt{intfloat/e5-large-v2}, \texttt{thenlper/gte-large}, \texttt{sentence-transformers/all-MiniLM-L6-v2}.

\section{Security Effectiveness}\label{sec:sec}
\textbf{Benign floor (no attacks).} Curated in-domain passages are statistically indistinguishable from
random (curated-vs-random area under the ROC curve, AUROC, $0.518$): genuinely-general documents are not hub-like; the cross-domain
Wikipedia set is below random ($0.414$). Organic natural hubs are separable from random
(AUROC $0.992$): they genuinely behave like hubs. Table~\ref{tab:floor} reports the three gate variants. With the threshold frozen on the disjoint calibration set (Section~\ref{sec:method}), the false-positive rate splits sharply by population (Table~\ref{tab:fpr}): genuinely-general documents are admitted cleanly (curated $1.1\%$, random $1.2\%$, Wikipedia $0.0\%$), but \emph{organic natural hubs are flagged at $86\%$}. This is not a tuning artefact: an organic natural hub and an adversarial hub are both hub-like, so the statistic alone cannot separate them. The gate therefore admits genuinely-general content cleanly and detects all universal and concept hubs at the pre-specified decisive operating point (recall $1.0$); organic natural hubs fall in the flagged set and constitute the residual that provenance (Section~\ref{sec:residual}) must adjudicate.

\begin{table*}[t]\centering
\caption{Benign false-positive floor (no attacks). Scores: global is a rate; per-topic and rknn-$z$ are
robust $z$-scores.}\label{tab:floor}
\begin{tabular}{lccc}
\toprule
Gate & N-organic p50/p99 & N-curated p99 & organic-vs-random AUROC\\
\midrule
global & 0.0011 / 0.0029 & 0.0007 & 0.992\\
per-topic & 3.0 / 9.0 & 3.0 & 0.965\\
rknn-$z$ & 6.0 / 16.0 & 4.0 & 0.992\\
\bottomrule
\end{tabular}
\end{table*}

\begin{table}[t]\centering
\caption{Leakage-free, leave-one-out benign false-positive rate per population: the operating threshold is frozen on the disjoint $5{,}000$-document calibration set (global gate, $\theta=0.0007$); corpus-derived candidates are scored pre-insertion (leave-one-out). Genuinely-general documents are admitted cleanly; organic natural hubs are not separable from adversarial hubs by the statistic.}\label{tab:fpr}
\begin{tabular}{lrrr}
\toprule
benign population & $N$ & FP & FPR\\
\midrule
organic natural hubs (top-$1\%$ rev-$k$) & 1{,}000 & 862 & 0.862\\
curated (definitional/overview) & 1{,}500 & 16 & 0.011\\
random corpus documents & 2{,}000 & 24 & 0.012\\
cross-domain Wikipedia & 1{,}500 & 0 & 0.000\\
\midrule
\multicolumn{4}{l}{\footnotesize attack recall at the same frozen threshold: universal $1.000$, concept $1.000$}\\
\bottomrule
\end{tabular}
\end{table}

\begin{figure}[t]\centering
\includegraphics[width=0.82\linewidth]{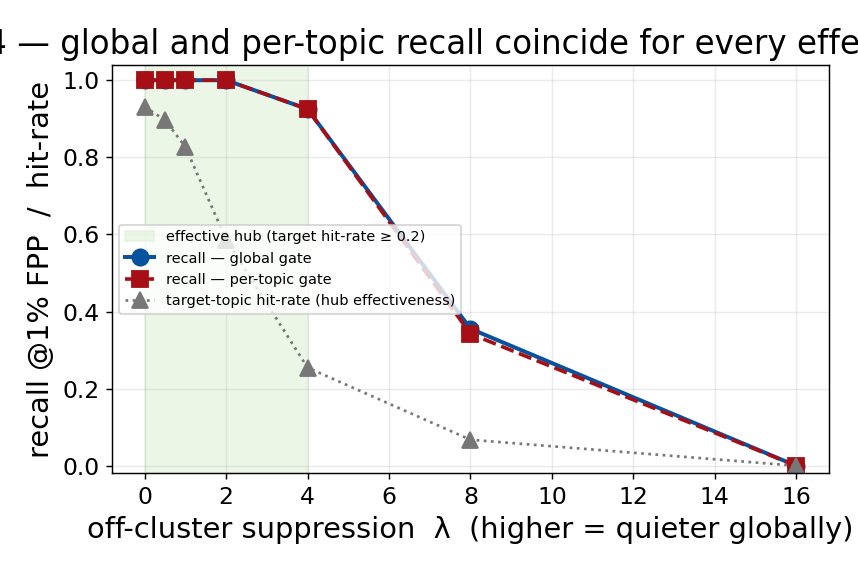}
\caption{Global and per-topic recall coincide for every effective hub; per-topic never exceeds global and
falls below once the attack is ineffective. Target hit-rate declines with the global hub rate as
$\lambda$ grows---the coupling predicted by anisotropy.}\label{fig:lambda}
\end{figure}

\textbf{Global vs.\ domain-aware (embedding-space attacks).} Across the suppression sweep
(Table~\ref{tab:lambda}, Fig.~\ref{fig:lambda}) per-topic recall equals global recall for every effective
hub and never exceeds it; the maximum marginal is $+0.000$. At $\lambda=8$ the hub already retrieves only
$7\%$ of its target topic---it has ceased to be effective---and there per-topic ($0.342$) is below global
($0.355$). On the embedding-space attacks the global gate reaches recall $1.0$ on universal hubs and on concept hubs at the decisive operating point $\lambda^{*}$ (AUROC $1.0$; $\ge 0.92$ across the effective $\lambda$ range, Table~\ref{tab:lambda}).

\begin{table*}[t]\centering
\caption{Gradient concept attack across the off-cluster suppression sweep $\lambda$ (embedding-space). Concept recall is $1.0$ at the pre-specified decisive point $\lambda^{*}{=}2$; the minimum over the effective range ($\lambda\le4$, target hit-rate $\ge20\%$) is $0.925$.}
\label{tab:lambda}
\begin{tabular}{lcccc}
\toprule
$\lambda$ & global rate & target hit-rate & recall global & recall per-topic\\
\midrule
0 & 0.211 & 0.93 & 1.000 & 1.000\\
2 ($\lambda^{*}$) & 0.066 & 0.59 & 1.000 & 1.000\\
4 & 0.011 & 0.25 & 0.925 & 0.925\\
8 & 0.000 & 0.07 & 0.355 & 0.342\\
16 & 0.000 & 0.00 & 0.000 & 0.000\\
\bottomrule
\end{tabular}
\end{table*}

\begin{figure*}[t]\centering
\includegraphics[width=0.92\linewidth]{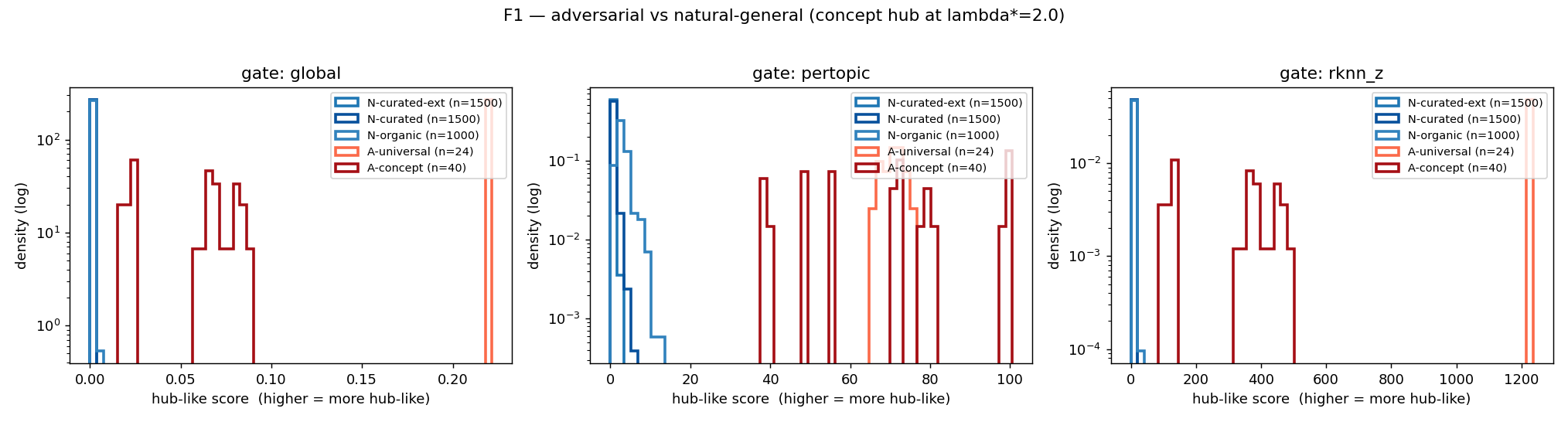}
\caption{Score distributions for natural-general negatives vs.\ universal and concept adversarial hubs.
Effective hubs separate cleanly; the natural-general docs sit at low scores. This panel is descriptive (score distributions); it is not evaluated at the frozen operating threshold---per-population false-positive rates at that threshold are in Table~\ref{tab:fpr}.}\label{fig:sep}
\end{figure*}

Fig.~\ref{fig:sep} shows the decisive score distributions: adversarial hubs separate cleanly from the
natural-general negatives under the global gate: adversarial hubs separate from genuinely-general documents, while naturally-occurring hubs overlap with the adversarial population and require provenance-based adjudication.

\textbf{Baselines.} On the same corpus, sentinels, and hubs the gate matches the reverse-$k$NN
detector~\citep{cisco2026ahd} at recall $1.0$/AUROC $1.0$ (Table~\ref{tab:cisco}); in our reproduction on this corpus both the global and domain-aware modes are recall-saturated, so this setting has no headroom to exhibit a domain-aware advantage; the detector's own paper reports settings where domain-scoped scanning recovers targeted attacks that evade global detection, consistent with our geometric account that such an advantage appears only where domains are isolated (the tight-domain residual of Section~\ref{sec:residual}). A provenance-only baseline flags every injected hub but also every untrusted natural document, so a provenance-only policy that rejects every untrusted document meets a $1\%$ benign false-positive rate only if at least $99\%$ of benign sources are already classified as trusted.

\begin{table*}[t]\centering
\caption{Reverse-$k$NN detector baseline~\citep{cisco2026ahd} on the same data.}\label{tab:cisco}
\begin{tabular}{llcccc}
\toprule
variant & family & r@0.4\% & r@1\% & AUROC & separation\\
\midrule
global & universal & 1.000 & 1.000 & 1.000 & 12.00\\
global & concept & 1.000 & 1.000 & 1.000 & 11.56\\
concept-aware & universal & 1.000 & 1.000 & 1.000 & 6.22\\
concept-aware & concept & 1.000 & 1.000 & 1.000 & 6.22\\
\bottomrule
\end{tabular}
\end{table*}

\textbf{Realised (HotFlip) attacks.} HotFlip-through-BGE hubs reach mean cosine $0.864$ ($0.84$--$0.89$)
of the embedding-space optimum (valid but non-fluent tokens). On the pre-specified run the global gate
recovers them at recall $1.0$ (AUROC $1.0$) and the per-topic gate at $0.93$ (AUROC $0.997$; this pre-specified seed is not bitwise-reproducible across machines, so we report the five-seed reproduction value of Section~\ref{sec:repro}), at a $1.1\%$ curated-document false-positive rate at the frozen operational threshold. The realised attack is the hardest case for both gates and is
analysed across seeds in Section~\ref{sec:repro}.

\begin{figure}[t]\centering
\includegraphics[width=0.74\linewidth]{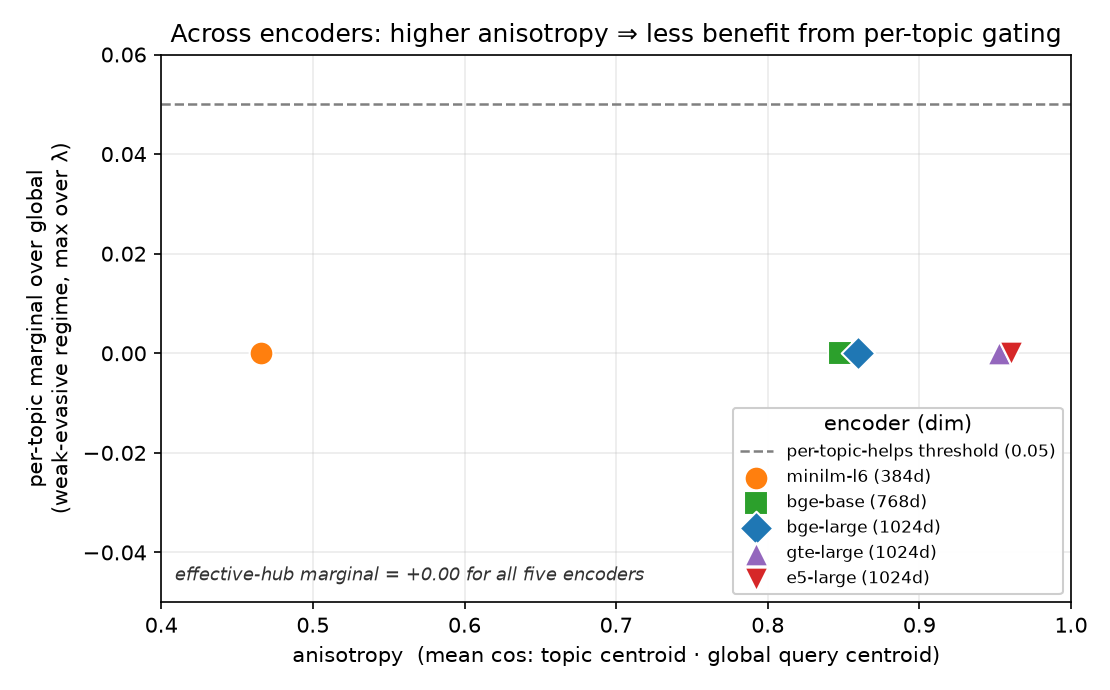}
\caption{Across five encoders the per-topic gate's marginal recall over the global gate is $\approx 0$
regardless of anisotropy: domain-aware refinement adds nothing on dense single-vector retrieval.}
\label{fig:enc}
\end{figure}

\textbf{Robustness to encoder.} Table~\ref{tab:enc} repeats the comparison on five encoders spanning
$384$--$1024$ dimensions and four architectures. The per-topic marginal over global is $+0.000$ on every
encoder; the global gate's universal recall is stable throughout (concept recall
is $1.0$ except on the most anisotropic encoder, e5-large, at $0.58$---a case that is hard for
\emph{both} gates). On the most isotropic encoder (MiniLM), the per-topic gate's curated-document FPR rises to $0.33$ under MiniLM (Table~\ref{tab:enc}) and its calibration is unstable across clusterings while the global gate remains stable---another reason to prefer
global. Anisotropy is reported as the mean cosine between per-topic query centroids and the global query
centroid; the per-topic marginal does not vary with it (Fig.~\ref{fig:enc}).

\begin{table*}[t]\centering
\caption{Per-topic gating adds nothing across encoders (full $N=10^5$). Anisotropy is the mean
topic-centroid$\cdot$global-centroid cosine; ``marginal'' is the maximum per-topic$-$global recall over
the $\lambda$ sweep.}\label{tab:enc}
\begin{tabular}{lccccc}
\toprule
encoder & dim & anisotropy & per-topic marginal & concept recall (global) & per-topic curated FPR\\
\midrule
all-MiniLM-L6-v2 & 384 & 0.466 & $+0.000$ & 1.000 & 0.330\\
bge-base-en-v1.5 & 768 & 0.848 & $+0.000$ & 1.000 & 0.000\\
bge-large-en-v1.5 & 1024 & 0.860 & $+0.000$ & 1.000 & 0.000\\
gte-large & 1024 & 0.952 & $+0.000$ & 1.000 & 0.000\\
e5-large-v2 & 1024 & 0.961 & $+0.000$ & 0.580 & 0.000\\
\bottomrule
\end{tabular}
\end{table*}

\textbf{Robustness to corpus.} To test generality beyond the assembled finance/biomedical corpus, we
rebuilt the entire pipeline on a compositionally distinct $100{,}000$-document general-web/\allowbreak encyclopedic
corpus (NQ, HotpotQA, DBpedia-entity, Quora; $37{,}033$ grounded queries; $8{,}000$ sentinels selected as in Section~\ref{sec:method}---the eval half of each domain's queries, capped at $7{,}500$, plus the $500$ cluster centroids). The
finding reproduces exactly (Table~\ref{tab:corpus2}): the benign floor holds (curated-vs-random AUROC
$0.514$, organic-vs-random $0.996$), the global gate recovers every effective concept hub at recall
$1.0$ (AUROC $0.999$) (curated-document FPR $0.9\%$, organic-hub flag rate $94\%$, mirroring corpus~1), and the per-topic marginal over global is again
$+0.000$ at \emph{every} suppression level. On this corpus effective hubs persist to higher $\lambda$
(the decisive hub at $\lambda^{*}=8$ still retrieves $41\%$ of its target topic and is caught at recall
$1.0$), so the global gate's sufficiency is, if anything, more pronounced. The negative result is thus
robust across two corpora of different domain composition in addition to the five encoders above.

\begin{table*}[t]\centering
\caption{The result reproduces on a second, compositionally distinct corpus (embedding-space attacks,
$N=10^5$, bge-large). Per-topic gating again adds nothing.}\label{tab:corpus2}
\begin{tabular}{lcc}
\toprule
metric & corpus 1 (finance/biomed) & corpus 2 (general-web)\\
\midrule
curated-document FPR & 0.011 & 0.009\\
random-document FPR & 0.012 & 0.011\\
organic-hub flag rate & 0.862 & 0.936\\
universal recall & 1.000 & 1.000\\
concept recall at $\lambda^{*}$ & 1.000 & 1.000\\
min.\ recall across effective $\lambda$ & 0.925 & 1.000\\
\textbf{per-topic marginal over global} & $+0.000$ & $+0.000$\\
\bottomrule
\end{tabular}
\end{table*}

\section{Geometric Analysis}\label{sec:geo}
The negative result is consistent with an anisotropy-based geometric mechanism (we show alignment, not sole causation). In these embedding spaces, per-topic query centroids are
positively aligned with the global query centroid (mean topic-centroid$\cdot$global cosine $0.47$--$0.96$; strong in four of five encoders), so the component of a candidate that raises its topic hub rate also raises its global
hub rate: a hub cannot be topic-loud and globally-quiet at once. A direct vector-space optimisation over the unit
sphere---maximising topic hub rate subject to a global-rate ceiling---did not find a topic-loud/global-quiet vector in the evaluated searches, and the target-topic hit-rate falls in lock-step with the global rate as $\lambda$ increases
(Table~\ref{tab:lambda}). This anisotropy is the well-documented narrow-cone geometry of contextual and
sentence embeddings~\citep{ethayarajh2019anisotropy,gao2019degeneration,mu2018allbut,li2020bertflow,gao2021simcse};
our result connects it to a security consequence: where the cone is narrow, a single global threshold is
sufficient and a domain-aware statistic is redundant.

\section{Systems Evaluation}\label{sec:systems}
We evaluate the gate end-to-end on a \emph{real} approximate index (HNSW via
hnswlib~\citep{malkov2018hnsw}) on the real \texttt{bge-large} embeddings, measuring the cost of running
admission synchronously on the ingestion path.

\textbf{End-to-end ingestion.} Per-document ingestion has four synchronous stages: encode $26.4$\,ms, candidate scoring $0.18$\,ms, threshold maintenance $0.72$\,ms (incremental $\tau$ update at the deployed $M{=}5{,}571$; p95 $1.1$, p99 $1.4$\,ms), and HNSW insert $2.0$\,ms (encode/scoring/insert p99 $51.6$/$0.32$/$3.8$\,ms)---a full ingestion path of $29.3$\,ms. The \emph{admission-control overhead}---candidate scoring plus synchronous threshold maintenance---is $(0.18{+}0.72)/29.3\approx3.1\%$, of which candidate scoring alone is $0.6\%$. The four stages are timed independently (scoring: the sentinel matmul and threshold compare; maintenance: the buffer update; insertion: the HNSW \texttt{add}) and reported as non-overlapping medians, so the $3.1\%$ does not double-count. Bulk index construction (hnswlib's multi-threaded \texttt{add\_items}) runs at $6{,}900$ docs/s and
adds $440$\,MB for $100$k vectors.

\begin{figure*}[t]\centering
\includegraphics[width=0.98\linewidth]{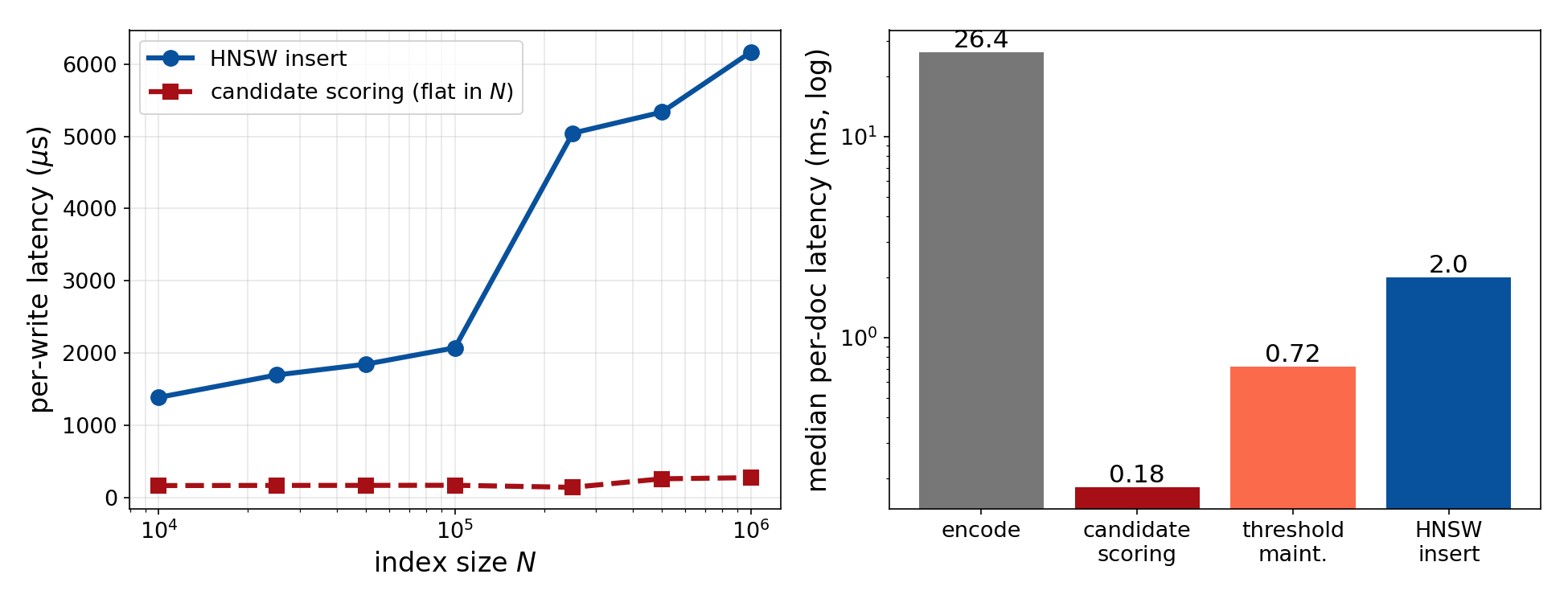}
\caption{Real-HNSW systems evaluation. Left: candidate-scoring latency remains flat through $N=10^6$, while HNSW insertion grows with index size. Right: median ingestion latency by stage (encode, candidate scoring, threshold maintenance, HNSW insert)---admission control adds $\approx3.1\%$ to median ingestion latency; candidate scoring alone contributes $0.6\%$.}\label{fig:sys}
\end{figure*}

\textbf{Scalability.} We separate two claims. \emph{Exact} incremental threshold maintenance was evaluated through $N=8\times10^4$ (Section~\ref{sec:sys}, Fig.~\ref{fig:incr}). Candidate-scoring and HNSW-insert latency were probed on an index grown to $N=10^6$---real corpus vectors through $N=10^5$, then synthetically extended for insert-timing only (Fig.~\ref{fig:sys}, left); the gate's per-write scoring latency is \emph{flat}
($139$--$273\,\mu$s at every $N$), while HNSW
insertion grows sub-linearly ($1.4$\,ms$\rightarrow$$6.2$\,ms). The gate's $N$-independence---the design's
central claim---holds on a real index, not just in the data-structure model of Section~\ref{sec:sys}.
Peak resident memory at $N=10^6$ is $9$\,GB.

\textbf{Approximate-index robustness.} The right question is not how far $\tau$ drifts but whether the admit/quarantine \emph{decision} changes when $\tau$ is maintained against a real HNSW index instead of exact $k$NN. With HNSW at recall@$10\approx0.93$--$0.96$, only $1.2\%$ of decisions flip over all candidates, and \emph{no} attack decision flips (universal and concept recall stay $1.0$); the flips concentrate in the borderline organic-hub population ($58/1000$, $5.8\%$); curated ($5/1500$), random ($8/2000$), and Wikipedia ($1/1500$) populations flip $<0.5\%$ and the attack families not at all ($0/24$ universal, $0/40$ concept) (underlying threshold drift $0.0015$ mean, p99 $0.017$). The gate's decisions are robust to the approximate index that production systems actually run.

\begin{figure*}[t]\centering
\includegraphics[width=0.98\linewidth]{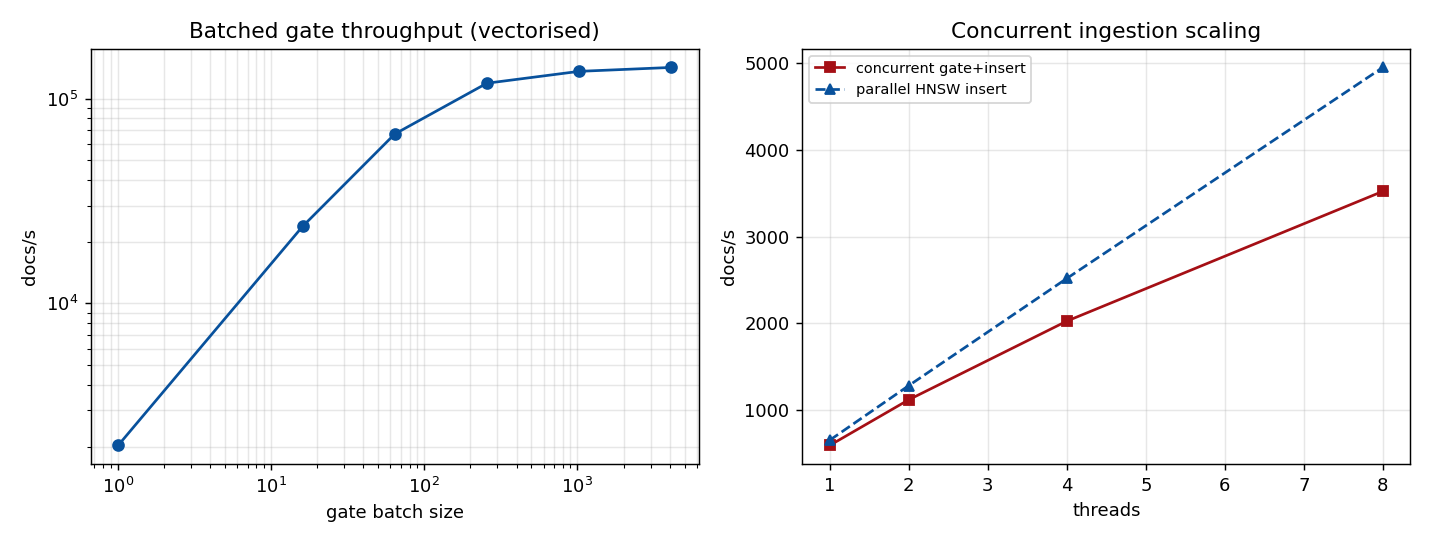}
\caption{Left: batched gate throughput (vectorised matmul). Right: concurrent ingestion scales with
threads; HNSW insertion remains the dominant operation, although admission control introduces measurable contention at higher concurrency.}\label{fig:conc}
\end{figure*}

\textbf{Concurrency and throughput.} The gate is a matmul and vectorises: batched scoring reaches
$142{,}000$ docs/s; HNSW insertion remains the dominant throughput bottleneck, though concurrent gate execution introduces measurable contention at higher thread counts. Under concurrent ingestion (Python
threads, each gating and inserting into a shared index), throughput scales $5.9\times$ across eight threads to $3{,}523$ docs/s---a $\approx29\%$ reduction versus insert-only at eight threads ($4{,}951\rightarrow3{,}523$ docs/s), attributable to Python-level contention and shared-index memory traffic (Fig.~\ref{fig:conc}).

\begin{figure}[t]\centering
\includegraphics[width=0.74\linewidth]{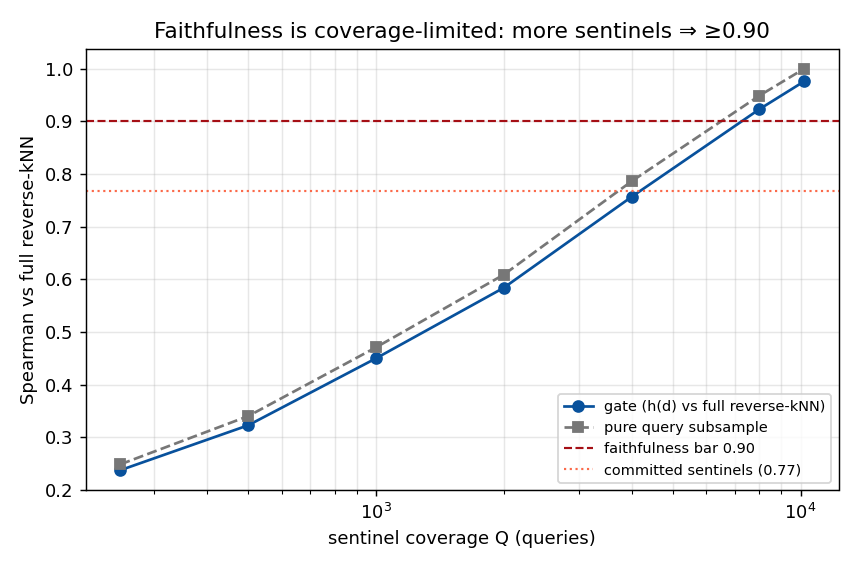}
\caption{Gate faithfulness is coverage-limited: Spearman against the full reverse-$k$NN scan rises with
sentinel coverage $Q$, reaching $\ge 0.90$ by $Q\!\approx\!8{,}000$ and $0.98$ at full query coverage.}
\label{fig:faith}
\end{figure}

\textbf{Faithfulness of the sentinel statistic.} The gate approximates a full reverse-$k$NN scan by a
sentinel sample; faithfulness is the Spearman correlation between $h(d)$ and the full reverse-$k$NN count.
With the evaluation configuration (sentinels drawn from the eval \emph{half} of queries, $\approx 50\%$
coverage) it is $0.77$ (Spearman over a $20{,}000$-document sample; a smaller $4{,}000$-document sample gives $0.75$); this is a coverage artefact, not a mechanism limit. Sweeping sentinel coverage
$Q$ (Fig.~\ref{fig:faith}) raises it monotonically to $0.92$ at $Q=8{,}000$ and $0.98$ at full query
coverage (a pure subsample reaches $1.000$, so the $\tau$ mechanism itself loses almost nothing). Because
candidate-scoring latency scales with $M$ and is $0.18$\,ms at $M=5{,}571$; it remains below $0.5$\,ms at full query coverage. Total admission overhead, including threshold maintenance (which also grows with the sentinel count), should be considered separately.

\section{Tight-Domain Residual and Complementary Provenance}\label{sec:residual}
Defining concepts as real corpus \emph{domains} (rather than $k$-means clusters) reproduces the
$+0.000$ per-topic marginal. The exception bounds the central claim: in a tight, distinct,
under-sampled domain (TREC-COVID, $50$ queries) an effective concept hub can be globally quiet and evade
\emph{both} gates, and there the concept-aware variant is \emph{worse} than global because natural
in-domain documents are themselves concentrated, raising the within-domain floor. This is consistent with
the geometric account: an isolated domain is less collinear with the global centroid. For such residual
cases, document \emph{provenance} (source trust) is a promising complementary control, orthogonal to the
geometric signal; we report a provenance-\emph{only} baseline and do not claim an integrated
gate-plus-provenance policy as an implemented contribution.

\section{Reproducibility}\label{sec:repro}
\begin{figure}[t]\centering
\includegraphics[width=0.74\linewidth]{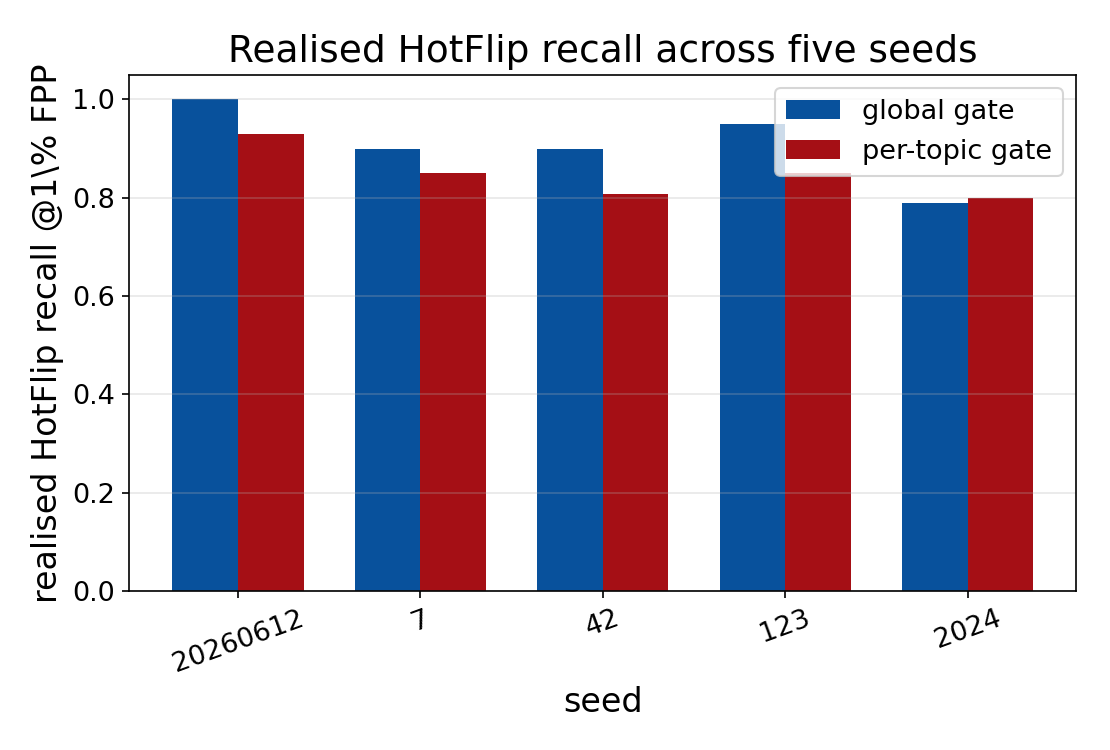}
\caption{Five-seed distribution of the realised-attack recall: per-topic never reliably exceeds global,
and the pre-specified seed is the most favourable for the per-topic gate.}\label{fig:seeds}
\end{figure}

The comparison between the global and domain-aware gates rests on two metrics near a decision boundary
and sensitive to nondeterminism in clustering, the anchor split, and attack construction. Re-running the
downstream pipeline for five seeds (reusing the cached corpus embedding) gives, on the realised HotFlip
attack, global recall $0.91\pm0.07$, per-topic recall $0.85\pm0.05$, and per-topic marginal over global
$+0.018\pm0.025$ (Tables~\ref{tab:seeds},~\ref{tab:seedsmarg}, Fig.~\ref{fig:seeds}): no statistically reliable improvement
from domain-aware refinement. The core separation is stable (universal recall $1.000$, embedding-space concept recall $1.0$ at the pre-specified decisive point (minimum recall $0.925$ across the effective suppression range), a $1.1\%$ curated-document false-positive rate at the frozen threshold, attack realisability
$\approx 0.86$). The pre-specified configuration is the most favourable of the five for the per-topic
gate, so we report the verdict conservatively: a single favourable run should not be read as a benefit.
A paired analysis (experimental unit: seed, $n=5$) of the realised-attack (per-topic $-$ global) recall
gives a mean of $-0.061$ ($95\%$ bootstrap CI $[-0.091,-0.023]$; Cohen's $d=-1.37$): per-topic is, if
anything, \emph{worse}, and the interval excludes zero on the side that favours global. The
embedding-space sweep marginal is $+0.018$ ($95\%$ bootstrap CI $[0.000,0.044]$, including zero). A
two-sided sign test is underpowered at $n=5$ ($p=0.38$); every interval is nonetheless consistent with no
benefit from domain-aware refinement. The full pipeline, frozen thresholds, and per-seed artefacts support full replication and are available from the author upon request.

\begin{table*}[t]\centering
\caption{Five-seed sweep, realised (HotFlip) attack: per-seed recall and AUROC. The pre-specified seed is the most favourable for the per-topic gate. SDs are population SD ($\mathrm{ddof}{=}0$).}
\label{tab:seeds}
\begin{tabular}{lccc}
\toprule
seed & global recall & per-topic recall & AUROC$_{pt}$\\
\midrule
20260612 (pre-spec.) & 1.000 & 0.930 & 0.997\\
7 & 0.900 & 0.850 & 0.975\\
42 & 0.900 & 0.807 & 0.995\\
123 & 0.950 & 0.850 & 0.993\\
2024 & 0.790 & 0.800 & 0.979\\
\midrule
mean$\pm$std & $0.910\pm0.070$ & $0.847\pm0.046$ & $0.988\pm0.009$\\
\bottomrule
\end{tabular}
\end{table*}

\begin{table}[t]\centering
\caption{Embedding-space suppression-sweep marginal (per-topic $-$ global, max over $\lambda$) by seed---a distinct quantity from the realised-attack recalls of Table~\ref{tab:seeds}.}
\label{tab:seedsmarg}
\begin{tabular}{lc}
\toprule
seed & emb-space sweep marginal\\
\midrule
20260612 (pre-spec.) & $+0.065$\\
7 & $+0.000$\\
42 & $+0.000$\\
123 & $+0.000$\\
2024 & $+0.025$\\
\midrule
mean$\pm$std & $+0.018\pm0.025$\\
\bottomrule
\end{tabular}
\end{table}

\section{Related Work}\label{sec:related}
\textbf{Hubness.} The concentration of reverse-nearest-neighbour counts in high-dimensional data was
characterised by Radovanovi\'{c} et al.~\citep{radovanovic2010hubs} and mitigated by similarity
re-scaling~\citep{schnitzer2012local}; we use it as a security signal rather than a retrieval-quality one.
\textbf{RAG and retrieval poisoning.} RAG~\citep{lewis2020rag,guu2020realm} over dense
retrievers~\citep{karpukhin2020dpr,khattab2020colbert,izacard2022contriever} is vulnerable to corpus
poisoning~\citep{zhong2023corpuspoison}, knowledge-corruption attacks
(PoisonedRAG~\citep{zou2024poisonedrag}), SEO/embedding attacks
(GASLITE~\citep{bentov2025gaslite}), and adversarial hubness~\citep{zhang2024advhub}; adversarial text is
realised by gradient token attacks~\citep{ebrahimi2018hotflip,zou2023gcg}. Defences include certified
output robustness (RobustRAG~\citep{xiang2024robustrag}), reranking/graph
defences~\citep{xiang2025grada}, and reverse-$k$NN hubness detection~\citep{cisco2026ahd}; provenance and
source-trust approaches secure the channel rather than the statistic. We add a
preventive, admission-time complement with a negative result on domain-aware refinement.
\textbf{ANN systems.} Production stores use HNSW~\citep{malkov2018hnsw}, IVF-PQ~\citep{jegou2011pq},
FAISS~\citep{johnson2021faiss}, ScaNN~\citep{guo2020scann}, and DiskANN~\citep{subramanya2019diskann}; we
show admission is cheap relative to insertion in such an index.
\textbf{Embedding anisotropy.} Contextual and sentence embeddings occupy a narrow
cone~\citep{ethayarajh2019anisotropy,gao2019degeneration}, addressed by all-but-the-top~\citep{mu2018allbut},
whitening/flow~\citep{li2020bertflow,su2021whitening}, and contrastive
training~\citep{gao2021simcse}; encoders used here are BGE~\citep{xiao2023bge}, E5~\citep{wang2022e5},
GTE~\citep{li2023gte}, MiniLM/SBERT~\citep{reimers2019sbert,wang2020minilm}, benchmarked with
BEIR~\citep{thakur2021beir}/MTEB~\citep{muennighoff2023mteb}. We tie anisotropy to a security sufficiency
condition.
\textbf{Incremental maintenance and admission control.} Our threshold buffers are an instance of
incremental view maintenance~\citep{blakeley1986ivm,gupta1995maintenance}; admission control is classic
in systems~\citep{welsh2003overload}. \textbf{Data poisoning.} Training-time and web-scale
poisoning~\citep{biggio2012poisoning,carlini2023poisoning} target learning; we target the retrieval
channel at write time.

\section{Limitations and Discussion}\label{sec:limits}
\textbf{Realised-attack recall.} On HotFlip the global gate is $0.91\pm0.07$, not perfect, and dips to
$0.79$ on one seed; both gates leave a realised-attack residual. \textbf{Tight-domain residual.} An
effective hub in an isolated, under-sampled domain can evade both gates (Section~\ref{sec:residual}); a
controlled domain-compactness/size sweep and an integrated provenance policy are future work.
\textbf{Encoder dependence.} Concept-hub separation is encoder-dependent (e5-large at $0.58$); the
\emph{per-topic-adds-nothing} result is robust across encoders, but absolute separation is not uniform.
\textbf{Single node.} The systems study measures one machine; multi-node/sharded distributed ingestion is
future work, as is end-to-end evaluation against live indirect-prompt-injection payloads. \textbf{Attack
scope.} Targeted single-query poisoning, multi-vector retrieval, and adaptive low-amplitude/multi-document
attacks are out of scope. \textbf{Hardware.} All systems measurements use an Apple-silicon (arm64) laptop with the MPS backend; the absolute latency and throughput figures should be re-validated on x86 server hardware with a discrete GPU, where the gate's $N$-independence and small relative overhead are expected to hold but the absolute numbers will differ. \textbf{Statistical power.} The paired bootstrap and effect size reported in
Section~\ref{sec:repro} rest on $n=5$ seeds; more seeds---particularly on the realised (HotFlip) attack---and more encoders would tighten the interval,
though the point estimates already favour the global gate.

\section{Conclusion}\label{sec:conc}
A preventive, admission-time \emph{global} statistical gate separates effective adversarial hubs from genuinely-general documents at a low measured false-positive rate (curated $1.1\%$, random $1.2\%$, Wikipedia $0.0\%$), in a realistic evasion setting and on par with a recent detector; naturally-occurring hubs remain statistically ambiguous and require provenance-based adjudication. The gate is maintainable incrementally at a per-write cost that is exact and $N$-independent (amortised for deletion), and---on a real HNSW index---adds $\approx3.1\%$ to the ingestion path (candidate scoring $0.6\%$), with $1.2\%$ of admit/quarantine decisions flipping under approximate indexing (none of them attacks). A domain-aware refinement provides no benefit in any evaluated setting, a result consistent with an anisotropy-based mechanism in which topic-local and global visibility are coupled. We advocate a global admission gate with provenance as a complementary control
for the tight-domain residual, and identify multi-node scalability and adaptive-adversary evaluation as
the primary next steps.

\section*{Declarations}
\textbf{Competing interests.} The authors declare that they have no known competing financial interests or personal relationships that could have appeared to influence the work reported in this paper.
\textbf{Funding.} This research received no external funding.
\textbf{Data availability.} The code and data supporting this study are available from the authors upon reasonable request.
\textbf{CRediT author contributions.} \textbf{Prashant Kumar Pathak:} Conceptualization, Methodology, Software, Validation, Formal analysis, Visualization, Writing -- original draft, Writing -- review and editing. \textbf{Tarun Kumar Sharma:} Conceptualization, Methodology, Validation, Writing -- review and editing.

\section*{Declaration of generative AI and AI-assisted technologies in the manuscript preparation process}
LLMs were used for editorial purposes in this manuscript, and all outputs were inspected by the authors to ensure accuracy and originality. The construction, experiments, and reported numbers are the authors' own work; the LLM did not run experiments or generate findings. No proprietary data was shared with the LLM, and the authors take responsibility for the contents of this paper.

\end{document}